\documentstyle[12pt,amsmath,amsfonts,amssymb]{article}
\newcommand{\arccot}{\mathop{\rm arccot}\nolimits}

\makeatletter
\renewcommand{\section}{\@startsection{section}{1}{0pt}%
{3.5ex plus 1ex minus .2ex}{2.3ex plus .2ex}%
{\large\bf}}
\makeatother
\author{A.V. Golovnev 
\qquad L.V. Prokhorov \\  \\
\quad {\small alex@amber.ff.phys.spbu.ru \qquad lev.prokhorov@pobox.spbu.ru}\\
{\small \it Saint-Petersburg State University, Saint-Petersburg, Russia}}
\title{UNCERTAINTY RELATIONS\\ IN CURVED SPACES}
\date{ }
\begin{document}
\maketitle
\abstract
Uncertainty relations for particle motion in curved spaces are discussed.
The relations are shown to be topologically invariant.
A new coordinate system on a sphere appropriate to  the problem is proposed. 
The case of a sphere  is considered in detail.
The investigation can be of interest for string and brane theory, solid state 
physics (quantum wires) and quantum optics.
\newpage
\section{Introduction}
The  Heisenberg uncertainty relation 
\begin{equation}
\label{heisenberg}
\Delta x \cdot \Delta p_x \geqslant \frac{\hbar}{2}
\end{equation}                                                                  
holds for quantum motion on a plane (see, for example, [1]);
here  $\Delta x$, $\Delta p_x$ are coordinate and momentum dispersions
respectively.
This inequality can be derived from the well known relation 
$\left [\hat x\,{,}\,\hat p_x\right ]=i\hbar$. However, in  quantum 
mechanics on a circle one has the standard commutational relation for
coordinate $\hat\varphi${\,}, with 
$\varphi\in\left [0{,}2\pi\right )$, and momentum  $\hat p_{\varphi}$,
 but the uncertainty relation cannot be stronger than 
 \begin{equation}
\label{ours}
\Delta\varphi\cdot\Delta p_{\varphi}\geqslant 0.
\end{equation}                                                                       
The momentum dispersion can be equal to zero while the coordinate one 
never becomes infinite.
In contrast to (\ref{heisenberg}),
inequality (\ref{ours}) \it is not informative at all,\rm\ 
since a product of two nonnegative values cannot be negative.
We have to mention here that in this paper an inequality of this type
will be regarded as an uncertainty relation only if it contains a
number in its rhs without dependence on the wave function. If we
neglect this requirement, stronger relations are possible (see, for
example, section 2) but they would be extremly sensitive to the choice
of coordinate system.

The problem holds, of course, for any compact manifold.
For the sphere this problem is even more complicated because of the absense of a
self-adjoint momentum operator related to the azimuth angle (due to 
 boundary terms in  matrix elements caused by the coordinate system poles).
We propose a solution of the problem for any coordinates with closed
coordinate lines.

We do not consider the peculiar properties of quantum mechanics connected with 
extrinsic and intrinsic geometries; they add an additional potential to
the Hamiltonian [2].

The foundations of quantum mechanics (see, for example, [3]) and, in 
particular, uncertainty relations [4,5] have been in the focus of 
unprecedented activity. Some subtle mathematical
points are taken into account, such as self-adjointness of operators,
their domains and so on [4,6], which are usually disregarded
in physical papers. 

Modern physics
often encounters quantum motion in curved spaces. In nanoelectronics
the quantum wires 
not only may be curved or closed, but sometimes fail even to be manifolds at all
(triple vertex, etc. [7,8]). Superstring theory [9] considers strings
and branes with different topologies. And, of course, our physical space
is obviously curved. Similar problems arise in quantum optics (photon number
and phase operators), but the main issue there is the correct definition of the
phase operator.

The cases of a circle and an arbitrary
curved line are presented in section 2.
Section 3 is devoted to quantum mechanics on the sphere; 
stereografic coordinates and  spaces  diffeomorphic to the plane are considered;
uncertainty relations are shown to be the usual ones. In section 4 we apply a
method of stereographic projection to the circle. New coordinates on 
the sphere, approapriate to the method of
section 2, are proposed in section 5.
 In section 6 and in the appendix we look at arbitrary manifolds
with closed coordinate lines; the diffeomorphic invariance of
uncertainty relations is shown. The phase space structure is discussed
in section 7.

\section{One-dimensional manifold} For the circle the situation described
above arises because the operator $\hat\varphi$ takes out physical states
from the Hilbert space of $2\pi$-periodical functions [4,6].
 Hence a product of operators $\hat\varphi$  
and $\hat p_{\varphi}$  is not well-defined and one should be very carefull
while working with these operators. It is easy to show that for any
normalized state $|\Psi\rangle$ the following inequality  is valid                      

\begin{equation}
\label{general}
\Delta\varphi\cdot\Delta p_{\varphi}\geqslant\left |Im\left\langle\hat\varphi%
\Psi |\hat p_{\varphi}\Psi\right\rangle\right |.
\end{equation}

Indeed, the definition of a dispersion for any observable reads
 [1,4,6,10]:
$\left (\Delta\varphi\right )^2=\left\langle\left (\hat\varphi -\bar%
\varphi\right )%
\Psi |\left (\hat\varphi -\bar\varphi\right )\Psi\right\rangle$, where $\bar%
\varphi$\ is a mean value of the observable. 
 Using the Cauchy inequality
$\left |\left\langle{\Psi}_1|{\Psi}_2\right\rangle\right |\leqslant|%
{\Psi}_1|\cdot%
 |{\Psi}_2|$,
 one gets [4,11]:
\begin{multline*}
\Delta\varphi\cdot\Delta p_{\varphi}\geqslant\left |%
\left\langle\left (\hat\varphi -\bar\varphi\right )\Psi |\left (\hat p_{%
\varphi} -\bar %
 p_{\varphi}\right )\Psi\right\rangle\right |\geqslant\\ \geqslant\left%
 |Im\left\langle%
\left (\hat%
\varphi -\bar\varphi\right )\Psi |\left (\hat p_{\varphi} -\bar p_{\varphi}%
\right )\Psi\right%
\rangle\right |=\left |Im\left\langle\hat\varphi\Psi |\hat p_{\varphi}\Psi%
\right\rangle\right |.
\end{multline*}  
 In this derivation we didn't use a product
of coordinate and momentum operators. If the product is well defined for any 
order of operators in it, then the rhs of (\ref{general})
 turns out to be $\frac12\left%
\langle\Psi%
|\left [\hat\varphi\,{,}\,\hat p_{\varphi}\right ]\Psi\right\rangle$, which
is the classical result.

On the circle one cannot use the commutator of 
$\hat\varphi$ and  $\hat p_{\varphi}$, but
 integration by parts and the $\left\langle%
\Psi|\Psi\right\rangle =1$ condition yield:
\begin{multline*}  
Im\left\langle\hat\varphi\Psi |\hat p_{\varphi}\Psi\right\rangle =%
Im\frac{\hbar}{i}\int_{0}^{2\pi}d\varphi\,\Psi^*(\varphi)\varphi%
\frac{\partial}{\partial\varphi}\Psi(\varphi)=\\
=-\frac{\hbar}{2}\int_{0}^{2\pi}d\varphi\,\varphi\frac{\partial}{%
\partial\varphi}\left | \Psi\right |^2=
\frac{\hbar}{2}\left (1-2\pi|\Psi(2\pi)|^2\right )=
\frac{\hbar}{2}\left (1-2\pi|\Psi(0)|^2\right ).
\end{multline*}
In general, if one rotates the coordinate system $\varphi\rightarrow%
\varphi+\delta \mod 2\pi$ the rhs of the obtained formula changes for a
given $\Psi$.  
For 
 ${\Psi}_k(\varphi)=\frac{1}{\sqrt{2\pi}}\exp\left (%
\frac{ik\varphi}{\hbar}\right )$ (here and hereafter $k$ is an integer
number times $\hbar$) the rhs of (\ref{general})
 equals zero, hence we have relation
 (\ref{ours}). It is the strongest possible relation with a 
pure number in its rhs.

This result can be obtained in another way. Let us consider a periodic
 coordinate operator $\Hat{\Tilde\varphi}%
 =f(\varphi)$,
demanding that $f(\varphi)=\varphi\mod 2\pi$ and $0\leqslant f(\varphi)<2\pi$
with the variable 
$\varphi\in(-\infty {,}\infty)$.
It defines the function properly and we have
\begin{equation}
\label{delta}  
\left [\Hat{\Tilde\varphi}\,{,}\,\hat p_{\varphi}\right ]=i\hbar\left (%
 1-\sum_{n=-\infty}^{\infty} 2\pi\delta(\varphi -2\pi n)\right ).
\end{equation}                    
The scalar product is $\left\langle\Psi_1|\Psi_2\right\rangle%
=\int\limits_{0-\epsilon}^{2\pi -\epsilon}\Psi_1^* \Psi_2 d\varphi$.
Only singularities at  $\varphi =0$  play a role while those at 
$\varphi =2\pi$ are disregarded. The standard relation 
$$\Delta\varphi\cdot\Delta p_{\varphi}\geqslant\frac12\left\langle\Psi |%
\left [\Hat{\Tilde\varphi}\,{,}\,\hat p_{\varphi} \right ]\Psi\right\rangle$$
and the inequality (\ref{general}) give the same result (\ref{ours}).

All this is valid for an arbitrary closed curve with a coordinate
 $\varphi%
\in\left [0{,}A\right )$  and a length element $dl=h(\varphi)d\varphi$ . We
 have to consider a self-adjoint operator $$\hat p_{\varphi}=\frac{\hbar}{i}%
\frac{1}{\sqrt{h(\varphi)}}\frac{\partial}{\partial\varphi}\sqrt{h(\varphi)}$$
and a scalar product $$\left\langle{\Psi_1}|{\Psi_2}\right\rangle =%
\int\limits_{0-\epsilon}^{A-\epsilon}{\Psi}_1^*{\Psi}_2h(\varphi)\,d\varphi.$$ 
One can refer to all the integrals as the ordinary ones and the rhs
of  (\ref{general}) can be integrated by parts. The 
$\delta$-function approach
 (\ref{delta})
also can be used. 
It yields again the previous result (\ref{ours}). The inequality is
saturated  by functions 
  $$\frac{1}{A\sqrt{h(\varphi)}}\exp \left (\frac{2i\pi k\varphi}%
{A\hbar}\right ).$$
\it So, the uncertainty relation (\ref{ours})
is invariant under any smooth deformation of the circle
 (topological invariance).\rm\  We
shall show that this property is valid not only 
for the one-dimensional case.

Note that due to the periodicity of all the functions
one can also consider only the upper limit singularities instead of the lower
 limit ones or take both with the factor $\frac12$ 
for each one.

\section{Stereographic projection}
As we show later, the methods of section 2 can be applied
to an arbitrary closed coordinate line on any manifold. But for the azimuth 
angle of spherical coordinates it does not work because the momentum 
operator is not self-adjoint (due to the boundary terms in coordinate
poles).

However in this case one may consider a stereographic projection.
It allows to introduce new operators on the projection plane:
\begin{equation}
\label{sphereone}
\left\{
\begin{aligned}
\hat q_1& =2R\cot\left (\frac{\vartheta}{2}\right )\cos\varphi,\\
\hat p_1& =\frac{i\hbar}{R}\left (\frac{\sin\varphi}%
{\sin\vartheta}%
\frac{\partial}{\partial\varphi}+\cos\varphi\frac{\partial}{\partial\vartheta}%
\right ){\sin}^2\frac{\vartheta}{2};
\end{aligned}
\right.
\end{equation}
\begin{equation}
\label{spheretwo}
\left\{
\begin{aligned}
\hat q_2& =2R\cot\left (\frac{\vartheta}{2}\right )\sin\varphi,\\
\hat p_2& =\frac{i\hbar}{R}\left (-\frac{\cos\varphi}%
{\sin\vartheta}%
\frac{\partial}{\partial\varphi}+\sin\varphi\frac{\partial}{\partial\vartheta}%
\right ){\sin}^2\frac{\vartheta}{2},
\end{aligned}
\right.
\end{equation}
with the lenght element $dl^2=R^2 (d{\vartheta}^2+{\sin}^2\vartheta\,d{\varphi}^2)=%
{\sin}^4\frac{\vartheta}{2}\left (dq_1^2+dq_2^2\right )$
 where
${\sin}^2\frac{\vartheta}{2}=\left (1+\frac{q_1^2+q_2^2}%
{4R^2}\right )^{-1}$,
 the surface area element  $$dS=R^2 \sin\vartheta d\vartheta d\varphi=%
{\sin}^4\frac{\vartheta}{2}\,dq_1 dq_2$$ and the scalar product
$\left\langle{\Psi}_1|{\Psi}_2\right\rangle=\int{\Psi}_1^*{\Psi}_2dS$.
Self-adjoint momentum operators on the sphere are given by
$$\frac{\hbar}{i}\frac{1}{{\sin}^2%
\frac{\vartheta}{2}}\frac{\partial}{\partial q_k}{\sin}^2%
\frac{\vartheta}{2}$$ 
 after rewriting the derivatives in terms of the spherical angles.
These operators are suggested by the  projection plane description.

Any differential operator can be written in terms of new coordinates and
momenta (\ref{sphereone}), (\ref{spheretwo}). For example, the free particle
kinetic energy is 
 $$-\frac{{\hbar}^2}{2}\Delta=\left (1+\frac{\hat q_1^2+\hat q_2^2}%
{4R^2}\right )\cdot%
\frac{\left (\hat p_1^2+%
\hat p_2^2\right )}{2}\cdot\left (1+\frac{\hat q_1^2+%
\hat q_2^2}%
{4R^2}\right ).$$

There are no periodic coordinates in this system and one has the standard
commutational and uncertainty relations $$\left [\hat q_1\,{,}\,%
\hat p_1\right ]=i\hbar,\qquad% 
\left [\hat q_2\,{,}\,\hat p_2\right ]=i\hbar,\qquad%
 \left [\hat p_1\,{,}\,%
\hat p_2%
\right ]%
=0,$$  
$$\Delta q_1\cdot\Delta p_1\geqslant\frac{\hbar}{2},\qquad%
 \Delta q_2\cdot\Delta p_2\geqslant%
\frac{\hbar}{2}.$$
The latter can be obtained using (\ref{general}) because the 
commutators are well defined at
a dense subset of physical wavefunctions.

Ordinary quantum mechanics on the plane is different due to fact that the area
element $dS=dx_1 dx_2$  does not have any factor tending to zero at the 
coordinate infinity which is drastic, of course, for
functions 
from the Hilbert space. Spherical 
infinity (the north pole) is an ordinary point
where a wavefunction can take any value 
while at the plane infinity wavefunctions must go to zero fast enough.
 It should be mentioned also that spherical coordinates
on the sphere are related to polar ones on the plane by the stereographic
projection; and polar coordinates have the same problem: the radial momentum
operator is not self-adjoint.

Note that all our formulae do not depend on the particular area element.
If one has a diffeomorphic image of the plane then the coordinate system
 $x_1$, $x_2$ can be moved to it, possibly with a different area element,
but with infinite points mapping onto infinite points.
Relation (\ref{heisenberg})
still holds, since the particular area element is irrelevant for our
 derivation.
 In other words,\ \it relation (\ref{heisenberg})
 is invariant under any smooth deformation of the plane.\rm\quad
Obviously, one can establish it also for an arbitrary manifold diffeomorphic
to the sphere by generalizing the operators
 (\ref{sphereone}) and  (\ref{spheretwo}). But one has to bear in mind that 
these coordinates can reach infinite values in spite of the finite total volume of 
the manifold. Another shortcoming is that some physical states
are excluded from the coordinate operator domains. We discuss it in more
detail in the next section.

Note also that the stereographic projection
is valid for an arbitrary dimension. For 
$S^n$ the lenght element is
$$dl^2={\sin}^4\frac{\vartheta}{2}\left (dq_1^2+dq_2^2%
+\ldots+dq_n^2\right ),$$
the surface element $dS={\sin}^{2n}\frac{\vartheta}{2}\,dq_1dq_2%
\cdots dq_n$,
 the momenta
$$\hat p_i=\frac{\hbar}{i}\frac{1}{{\sin}^n\frac{\vartheta}{2}}%
\frac{\partial}{\partial q_i}{\sin}^n\frac{\vartheta}{2},$$ 
the commutational and the uncertainty relations are standard.
The free particle kinetic energy is:
\begin{multline*}
-\frac{{\hbar}^2}{2}\Delta=\frac12{\left (1+\frac{\hat q_1^2+\ldots+%
\hat q_n^2}{4R^2}%
\right )}^{n/2}
\cdot\\   \cdot
\sum_{i=1}^{n}\left (\hat p_i {\left (1+\frac{\hat q_1^2+%
\ldots+\hat q_n^2}{4R^2}%
\right )}^{4-2n}\hat p_i\right )\cdot%
 {\left (1+\frac{\hat q_1^2+\ldots+\hat q_n^2}{4R^2}\right )}^{n/2}.  
\end{multline*}

\section{Stereographic projection for the circle} In section 3 
we obtained the standard
relations (\ref{heisenberg}) for the quantum motion on any n-dimensional
sphere, $n\geqslant2$. In the same manner one 
can get (\ref{heisenberg})  for $n=1$, but earlier we had
 (\ref{ours}) for the circle. What is the matter? To elucidate why we had 
different results for the circle, let us consider 
a stereographic projection from the circle onto the line with zero angle being
at the contact point and $\varphi\in\left [%
-\pi{,}\pi\right )$. Then $x=2R\,\tan\frac{\varphi}{2}$. Thus we use in this
section the $\left [-\pi{,}\pi\right )$ interval instead of the
$\left [0{,}2\pi\right )$ one. This trick changes the operator dispersions [12]
but, as can be easily seen, leaves relation (\ref{ours}) the same.

We start from the following generalization: $x\to\tilde x=%
\frac{2R}{\alpha}\tan\frac{\alpha\varphi}{2}$,
$\alpha\in[0{,}1]$. The lenght element is 
$dl=R\,d\varphi={\cos}^2\frac{\alpha\varphi}{2}\,d\tilde x$  and the momentum 
$$\hat p_{\tilde x}=\frac{\hbar}{iR}\cos\frac{\alpha\varphi}{2}%
\frac{\partial}{\partial\varphi}\cos\frac{\alpha\varphi}{2}$$  
(its self-adjointness is a matter of direct calculation). 
The $\tilde x$  domain
 is $\left [-\frac{2R}{\alpha}\tan\frac{\alpha\pi}{2}{,}%
\frac{2R}{\alpha}\tan\frac{\alpha\pi}{2}\right )$. If $\alpha\to 0$ one has
$\tilde x\to R\varphi$ and $\hat p_{\tilde x}\to\frac{\hbar}{iR}%
\frac{\partial}{\partial\varphi}$, i.e. ordinary quantum mechanics on the
circle with the angle variable $\varphi $ and
the uncertainty relation (\ref{ours}).
 If $\alpha\to 1$ then  $\tilde x\to x\in{\mathbb R}^1$ 
while  $$\hat p_{\tilde x}\to\frac{\hbar}{iR}\cos\frac{\varphi}{2}%
\frac{\partial}{\partial\varphi}\cos\frac{\varphi}{2},$$ so the standard relation
 (\ref{heisenberg}) is valid for
 $\tilde x$
and $p_{\tilde x}$. And what about arbitrary  $\alpha$\,?  A simple 
calculation yields for the rhs of (\ref{general}):
 \begin{multline}
\label{circle}
Im\left\langle\Hat{\Tilde x}\Psi |\hat p_{\tilde x}\Psi\right\rangle=\\
=Im\int\limits_{-\pi}^{\pi}d\varphi\,\frac{2R}{\alpha}\tan\frac{\alpha\varphi}{2}%
{\Psi}^*(\varphi)\frac{\hbar}{iR}\cos\frac{\alpha\varphi}{2}%
\frac{\partial}{\partial\varphi}\cos\frac{\alpha\varphi}{2}\Psi (\varphi)=\\
=\frac{\hbar}{2}\left (1-\frac{\sin\alpha\varphi}{\alpha}{\left.\left |%
\Psi (\varphi)%
\right |^2\,\right |}_{-\pi}^{\pi}\right )=\frac{\hbar}{2}\left (1-%
\frac{2\sin\alpha\pi}{\alpha}{\left |\Psi(-\pi)\right |}^2\right ),
\end{multline}                                   
where the periodicity $\Psi(-\pi)=\Psi(\pi)$ and the normalization condition
$\langle\Psi|\Psi\rangle=1$ were taken into account.
For any $\alpha <1$  the uncertainty relation has the form of 
 (\ref{ours}): $\Delta\tilde x\cdot\Delta%
 p_{\tilde x}\geqslant 0$  (it is saturated by any function with
 $\left |\Psi(\pi)\right |^2%
=\frac{\alpha}{2\sin\alpha\pi}$\ ). But at $\alpha=1$  we have
the standard relation
 (\ref{heisenberg}) since $\sin\alpha\pi\to 0$  as $\alpha\to 1$.
This is because the nasty point $\varphi =-\pi$ is moved away to infinity
and the dispersion of $\tilde x$  gains infinite values.
The single point $\alpha =1$ does not have any corresponding
 value of  $\tilde x$.
Taking this point out from the manifold changes the topology and 
uncertainty relations. But it changes nothing for smooth functions: 
this coordinate is good enough because we are short
of  one point only.

Now we have to discuss the domains of operators under consideration.
If $\alpha < 1$, all the functions are defined on a finite interval.
If we demand them to be equal to zero at the ends of this interval then
the momentum operator would be symmetric but not self-adjoint. It admits
an infinite number of self-adjoint extensions but only one of them,
$\Psi(-\pi)=\Psi(\pi)$, ensures the finiteness of energy. We use this extension
from the very beginning because it is natural for quantum mechanics on the
circle.

In the limit of $\alpha=1$\quad $\hat p_{\tilde x}$ is self-adjoint but its domain
contains wavefunctions of infinite energy states (those with a discontinuity
at $\varphi=\pm\pi$). Still, it can result only in some additional unphysical
states for which uncertainty relations are valid. It is more important that
$\Hat{\Tilde x}$ with $\alpha=1$ {\it does not admit some physical states}
(those with $\int\limits_{-\pi}^{\pi}\left|\Psi(\varphi)\right|^{2}{\tan}%
^{2}\frac{\varphi}{2}d\varphi=\infty$). In principle, it is not a problem due to
the following facts: 1) the matrix element (\ref{circle}) is correctly defined
for all the physical states and depends on them smoothly; 2) in the $L^2$
metrics any function $\Psi(\varphi)$ can be regarded as a limit of
functions $\Psi_{\omega}(\varphi)=\Psi(\varphi)\omega(\varphi)$ with
$\omega(\varphi)$ equal to unity everywhere apart from a small region around
$\varphi=\pm\pi$ and tending to zero fast enough while approaching this point;
3) the momentum dispersion for $\left|\Psi_{\omega}\right>$ tends to that for
$\left|\Psi\right>$ when $\omega(\varphi)\to 1$ due to the $\cos\frac{%
\varphi}{2}$ factors in the $\hat p_{\tilde x}$ definition. But nevertheless,
this problem exists for spheres of any dimensionality and it is one more
reason to search for better coordinates, which is done in the next section.
  
\section{Coordinates on sphere} One of the problems of quantum mechanics
on a sphere is a proper choice of canonical variables. Spherical
coordinates do not solve the problem because they do not provide
self-adjoint momenta. Stereographic coordinates are better but they are 
infinite in the north pole which is just an ordinary point
and their operators have too small domains. It results
in (\ref{heisenberg}). We overcome this difficulty by  ``wrapping''
coordinate lines from the projection plane onto the sphere. In the case of
two dimensions one can write:
\begin{equation}
\label{newco}
\left \{
\begin{aligned}
\eta& =2\arccot\left (\cot\frac{\vartheta}{2}\cos\varphi\right ),\\
\xi& =2\arccot\left (\cot\frac{\vartheta}{2}\sin\varphi\right ).
\end{aligned}
\right.
\end{equation}
Actually, this can be easily generalized to any other dimensionality.
Lines of constant $\eta$  or $\xi$ are loops "hanging down" from the north
 pole. New coordinates have finite ranges,
$\eta\in\left [0{,}2\pi\right )$, $\xi\in\left [0{,}2\pi\right )$,
hence we do not encounter the problems of the $\alpha=1$ case of section 4,
but the appropriate Hilbert space contains only $2\pi$-periodic functions
with respect both to $\eta$ and $\xi$.

One can check that thus we have the orthogonal coordinate system on the
sphere with closed coordinate lines and one singular point, the north pole.
 The length element and the metric tensor determinant are
\begin{multline*}
dl^2=R^2\left ({\sin}^2\frac{\vartheta}{2}+{\cos}^2\frac{\vartheta}{2}%
{\cos}^2\varphi\right )d{\eta}^2+\\
+R^2\left ({\sin}^2\frac{\vartheta}{2}+{\cos}^2\frac{\vartheta}{2}%
{\sin}^2\varphi\right )d{\xi}^2,
\end{multline*}
$$g=R^4{\left ({\sin}^2\frac{\vartheta}{2}+{\cos}^2\frac{\vartheta}{2}%
{\cos}^2\varphi\right )}^2\cdot{\left ({\sin}^2\frac{\vartheta}{2}%
+{\cos}^2\frac{\vartheta}{2}{\sin}^2\varphi\right )}^2.$$ 
In this formulae $\vartheta$ and $\varphi$ should be regarded as 
functions of new coordinates, $\eta$  and $\xi$. The surface element is 
$dS=g^{1/2}(\eta{,}\xi )d\eta %
d\xi$. The momentum operators
$$\left \{
\begin{aligned}
\hat p_{\eta}& =\frac{\hbar}{ig^{1/4}}\left (%
\frac{\partial\vartheta}{\partial\eta}\frac{\partial}{\partial\vartheta}+%
\frac{\partial\varphi}{\partial\eta}\frac{\partial}{\partial\varphi}%
\right )g^{1/4},\\
\hat p_{\xi}& =\frac{\hbar}{ig^{1/4}}\left (%
\frac{\partial\vartheta}{\partial\xi}\frac{\partial}{\partial\vartheta}+%
\frac{\partial\varphi}{\partial\xi}\frac{\partial}{\partial\varphi}%
\right )g^{1/4}
\end{aligned}
\right.$$
are written down in terms of old variables, $\vartheta$  and $\varphi$.
Commutational relations
$$\left [\hat\eta\,{,}\,\hat p_{\eta}\vphantom{\hat\xi}\right ]=i\hbar,%
\qquad\left [%
\hat\xi\,{,}\,\hat p_{\xi}\right ]=i\hbar,\qquad\left [%
\hat p_{\eta}%
\,{,}\,\hat p_{\xi}\vphantom{\hat\xi}\right ]=0$$  can be verified directly
but, as in section 2, one has to remember that the lhs of these equations
are not defined for the majority of states (it is again a question of
periodicity).

One may derive uncertainty relations in two different ways: either using the
 $\delta$-function approach and a relation analogous to
(\ref{delta}) or applying the inequality (\ref{general}).
 In any case the result is
 $$\Delta\eta%
\cdot\Delta p_{\eta}\geqslant 0,\qquad \Delta\xi\cdot\Delta p_{\xi}%
\geqslant 0.$$ It is saturated by
\begin{multline*}
{\Psi}_k\left (\vartheta{,}\varphi\right )=\frac{1}{Ag^{1/4}}\cdot\\
\cdot\exp\left (\frac{2ik_1}{\hbar}\arccot\left (\cot\frac{\vartheta}{2}%
\cos\varphi\right )+\frac{2ik_2}{\hbar}\arccot\left (\cot\frac{\vartheta}{2}%
\sin\varphi\right )\right ),
\end{multline*}
$A$ being a normalization constant.
Coordinates $\eta$ and $\xi$ are preferable because they 
have finite ranges and do not attribute infinite values to finite points.
In general, if any two points separated by a finite distance
have a finite difference in their coordinates  and all the 
distances on the manifold are limited by some constant then all coordinates 
will have finite domains. In particular,  every manifold with a positive
curvature greater than some positive constant is of this type [13].
If, in addition, all coordinate lines are closed, the inequality
 (\ref{ours}) is obtained.
 
\section{Topological invariance} If relation (\ref{ours}) holds
for some manifold with closed coordinate lines on it,
then  \it it holds also for any manifold diffeomorphic to the initial one
\rm\  with a coordinate system being moved from one manifold to another
by relative diffeomorphism. Disregarding for a moment the smoothness 
properties, 
one can speak about homeomorphisms instead of diffeomorphisms
and about topological invariance.

Indeed, 
the inequality (\ref{ours}) is valid for a coordinate system with finite coordinates
and closed coordinate lines. It is not changed under any diffeomorphism, hence
we only have to obtain relation (\ref{ours}) in general case. Actually,
we have already shown that for two-dimensional manifolds; now we shall 
do it for an arbitrary dimensionality and in some more detail.

Consider a manifold $M^n$ with a global orthogonal coordinate system.
Let one of the coordinates be $x$, and all the others $y$.
Suppose we have a finite domain and closed coordinate lines for $x$,
 $x\in\left [0{,}A(y)\right )$. Define a self-adjoint momentum operator
 $$\hat p_x=\frac{\hbar}{i}\frac{1}{\sqrt[4]{g(x,y)}}%
\frac{\partial}{\partial x}\sqrt[4]{g(x.y)},$$ where $g(x,y)$ is the metric
tensor determinant for the manifold. Then the rhs of
 (\ref{general}) yields
\begin{multline*}
Im\frac{\hbar}{i}\int\limits_{M^n}dxdy\sqrt[4]{g(x,y)}\,x\Psi (x,y)%
\frac{\partial}{\partial x}\sqrt[4]{g(x,y)}{\Psi}^*(x,y)=\\
=\frac{\hbar}{2}\int\limits_{M^n}dxdy\left (\sqrt{g(x,y)}{\left |\Psi%
 (x,y)\right |}^2-\frac{\partial}{\partial x}\left (\sqrt{g(x,y)}\,x%
{\left |\Psi (x,y)\right |}^2\right )\right )=\\
=\frac{\hbar}{2}\left (1-\int\limits_{M^n\cap\{x=0\}}dy\sqrt{g(0,y)}\,%
A(y){\left |\Psi (0,y)\right |}^2\right );
\end{multline*}
the normalization to unity is taken into account. 
The result equals to zero for functions
$${\Psi}_k(x,y)=\frac{\exp\left (\frac{2i\pi kx}{\hbar A(y)}\right )}%
{\sqrt{\int\limits_{M^n\cap\{x=0\}}dyA(y)}\,\sqrt[4]{g(x,y)}}.$$ 

Again we have the uncertainty relation
 (\ref{ours}):
$\Delta x\cdot\Delta p_x\geqslant 0$. It does not depend upon
  $g(x,y)$ and $A(y)$, hence it \it is invariant under diffeomorphisms
  \rm of $M^n$, as they do not change the topological structure,
but only vary the functions mentioned above. The proper coordinate systems 
can be found at least on the sphere (formulae (\ref{newco})) and on the
torus (obvious, since topologically
 ${\mathfrak T}^2\cong S^1\times S^1$), hence the result (\ref{ours})
is valid for any manifold diffeomorphic to the sphere or to the torus.

In general, the problem is to prove the existence of such a coordinate system.
We need it in order to 
define the operators under consideration but not all manifolds possess
such systems. Nevertheless, one 
coordinate can be defined if there is a smooth vector field on a manifold
with closed integral curves of finite lengths and not taking  zero values
except from a numerable set of singular points $\{t_m\}$, $m=1,2,3\ldots$ \ .
Any manifold homeomorphic to a direct product of a circle and an arbitrary 
manifold, regardless of how complex  it is, serves as an example. It has a 
vector field without singular points generated by rotations of the circle.
The vector field defines  coordinate and momentum operators, even if the
global coordinate system does not exist.
In appendix we show it in general case.

\section{Smooth manifolds and phase spaces} For a free particle on a
smooth manifold $M^n$, its phase space is the tangent bundle $TM^n$. This
bundle is always locally trivial [16] and within any given map allows
us to define n momenta (one momentum for each coordinate). But in general
it can not be made uniformly for the whole manifold because the manifold
neither has a global coordinate system nor is the bundle globally trivial
and admits a smooth section.

The above consideration is appropriate for the following two simplest
cases:

a) The manifold $M^n$ admits a global coordinate system, possibly, 
with closed coordinate lines, but without singular points. It means
that $M^n$ possesses a complete set of independent smooth vector fields
 (complete parallelizability; Euclidean spaces ${\mathbb R}^n$, tori
${\mathfrak T}^n\cong S^1\times\cdots\times S^1$, cylinders $S^1\times%
{\mathbb R}^n$, 1-, 3- and 7-dimesional spheres $S^1,\ S^3,\ S^7$, etc). In this case the tangent
bundle  is trivial and one can define global coordinates and momenta for the
whole $M^n$.

b) The manifold $M^n$ admits a global coordinate system with a numerable set
(actually, in the previous sections it was a one-element set) of singular
points (for example, $S^n\ \forall n$). In this case $TM^n$ becomes trivial
after taking away these points from the manifold. Hence one can define
global coordinates and momenta everywhere except from the numerable set
of points which means nothing for the smooth function properties. In
section 5 we have shown how it works.

If the $TM^n$'s structure is not so easy the situation is more intricate.
One cannot introduce a global coordinate system and the problem of
coordinate-momentum uncertainty relations may become senseless because
the operators under consideration simply do not exist. Still sometimes
it is possible to define m coordinates and momenta, $m<n$, even in
this case. For example, in the appendix a manifold with one coordinate
common for all the maps is investigated. Loosely speaking, $TM^n$ behaves
as a trivial bundle with respect to the common coordinate. A tangent space
can be moved along the coordinate line and its initial position is 
restored after completing a revolution.

In the general case
global coordinates are no longer defined and one has to use
other observables.

\section{Conclusion} Thus, for finite coordinates with closed coordinate 
lines the uncertainty relation has the form
 $\Delta x\cdot\Delta p_x\geqslant 0$, and it cannot be made stronger.
This is obvious for the eigenstates of the momentum operator. 
We have shown above how it can be reconciled with the canonical
 commutational relations. It is worth stressing that one has to make a
good choice of the coordinate system on a manifold for the self-adjoint
momenta to be well defined.

The uncertainty relation problem for the phase and the number of photons
in quantum optics is not directly related to the problem for
 $\hat p_{\varphi}$ and $\hat\varphi$. Creation and annihilation operators
${\hat a}^+$, $\hat a$ are not normal ones (they do not commute). Hence,  
they do not have a decomposition $\hat a=\sqrt{\hat a{\hat a}^+}\,\hat U=%
\hat U\sqrt{\hat a{\hat a}^+}$ with a unitary operator 
$\hat U$. The problem is to define the phase operator correctly [14,15].
\par
{\vspace*{2ex}
{\large\bf\begin{center} Appendix. \end{center}}
\vspace{2ex}}

Consider a manifold $M^n$ possessing a smooth vector field with closed integral
curves of finite length and a hypersurface orthogonal to the curves.
Let us attribute the zero value of the first coordinate ($x_1=0)$ to the latter.
The hypersurface points ($\eta$) parametrize the set of integral curves.
The first coordinate changes along the curves and can be made equal to
the path length along the curve from the zero surface in the fixed direction:
 $x_1\in\left [0{,}A(\eta)%
\right )$, with $A(\eta)$ being the integral curve length.

Let $\phi_{\alpha}:\,U_{\alpha}\to%
\phi_{\alpha}(U_{\alpha})$ be a set of maps on the manifold with
 $U_{\alpha}$ in $M^n$ and
 $\phi_{\alpha}(U_{\alpha})$ in Euclidean space
${\mathbb R}^n$. Choose any
$\epsilon >0$ and cut off balls  
 ${\mathbb B}_{{\epsilon}^m}(t_m)$ of radii 
${\epsilon}^m$ ($\epsilon$ is powered by m for total cut off volume to tend to zero as
$\epsilon\to 0$) around the numerable set of singular points and also cut off 
a vicinity of
 $x_1=0$ surface: 
$${{\tilde M}^n}_{\epsilon}=M^n \setminus\left (\left\{x_1\geqslant A\left (1-%
\epsilon\right )\right \}\cup\left \{x_1\leqslant\epsilon A\right\}\cup%
\left (\bigcup_m{\mathbb B}_{{\epsilon}^m}(t_m)\right )\right )$$
with new maps $\tilde U_{\alpha}=U_{\alpha}\cap{\tilde%
M}^n_{\epsilon}$  and $\tilde\phi_{\alpha}={\left.\phi_{\alpha}\right |}_{%
\tilde U_{\alpha}}$. The vector field can be moved [17] from $\tilde U_{\alpha}$
  to $\tilde\phi_{\alpha}\left (\tilde U_{\alpha}\right )$  and  the first
coordinate $x_1$ in the map can be put equal to the one
introduced invariantly above: ${\left (%
\tilde\phi_{\alpha}(\zeta)\right )}_1=x_1(\zeta),\quad\forall\zeta\in%
\tilde U_{\alpha}$. Now, let us choose an orthogonal coordinate system 
in every set $\tilde\phi_{\alpha}%
(\tilde U_{\alpha})$  with this first coordinate 
$x_1(\zeta)$. The corresponding operator can be defined 
invariantly and in maps as follows:
$$\hat s\Psi (\zeta)=x_1(\zeta)\Psi (\zeta)={\left (\tilde\phi_{\alpha}%
(\zeta)\right )}_1{\Psi}_{\alpha}^{\prime}\left (\tilde\phi_{\alpha}(\zeta)%
\right ),\quad\forall\zeta\in\tilde U_{\alpha},$$
where  ${\Psi}_{\alpha}^{\prime}=\Psi\circ\tilde\phi_{\alpha}^{-1}$.

The measure can be written as $d\mu=\left (\prod_{i}dx_{i}^{\alpha}%
\right )g_{\alpha}^{1/2}(x_{i}^{\alpha})$, where $x_{i}^{\alpha}$ are
coordinates in $\tilde\phi_{\alpha}(\tilde U_{\alpha})$. One can introduce
a momentum operator in this map $\hat p_{s}^{\alpha}=\frac{\hbar}{i}%
\frac{1}{g^{1/4}}\frac{\partial}{\partial x_1}g^{1/4}$
which generates a shift along an integral curve of the vector
field and
 is invariant under
any change of map (the Jacobian does not depend on $x_1$). It can be 
defined uniformly for the whole manifold
$\hat p_s\Psi(\zeta)=\sum\limits_{\alpha}{\chi}_{\alpha}(\zeta)%
\hat p_{s}^{\alpha}\Psi(\zeta)$
with ${\chi}_{\alpha}$ being a smooth partition of unity for $\{\tilde%
 U_{\alpha}{,}\tilde\phi_{\alpha}\}$. 

In the matrix element
$$\left\langle{\Psi}_1|\hat p_s{\Psi}_2\right\rangle=%
\lim\limits_{\epsilon\to 0}\int\limits_{{\tilde M}_{\epsilon}^{n}}d\mu\,%
{\Psi}_1^*\hat p_s{\Psi}_2$$ one can integrate by parts and all nonintegral 
terms vanish at the edges of maps with the partition of unity functions
${\chi}_{\alpha}$ 
except of cutoffs where they cancel each other when $\epsilon\to 0$ . Therefore
$$\left\langle{\Psi}_1|\hat p_s{\Psi}_2\right\rangle=\left\langle%
\hat p_s{\Psi}_1|%
{\Psi}_2\right\rangle-\frac{\hbar}{i}\sum\limits_{\alpha}\int d\mu%
\frac{\partial{\chi}_{\alpha}}{\partial x_1}{\Psi}_1^*{\Psi}_2=%
\left\langle\hat p_s{\Psi}_1|{\Psi}_2\right\rangle$$
because $\sum\limits_{\alpha}\chi_{\alpha}\equiv 1$.  We
have proved that
$\hat p_s$  is self-adjoint.

Consider now $\left\langle\hat s\Psi|\hat p_s\Psi\right\rangle$. 
If one integrates it by parts, the cancellation along the cutoff
 $x_1=0$ does not occur because  the coordinate $x_1$ is 
not periodic. Let 
${\alpha}^+$ be  such $\alpha$ that $\tilde U_{\alpha}$ touches  the edge
 $x_1=A\left (1-\epsilon\right )$, and ${\alpha}^-$ be  such
$\alpha$ that $\tilde U_{\alpha}$  touches the edge
 $x_1=\epsilon A$. Then (with the normalization condition):
\begin{multline*}
\frac{2}{\hbar}Im\left\langle\hat s\Psi|\hat p_s\Psi\right\rangle=%
\sum_{\alpha}\int\limits_{{\tilde M}_{\epsilon}^n}
dx_{1}^{\alpha}\cdots %
dx_{n}^{\alpha}g_{\alpha}^{1/2}(x^{\alpha}){\left |{\Psi}_{\alpha}%
(x^{\alpha})\right |}^2\\
-\sum_{{\alpha}^+}\int\limits_{{\tilde M}_{\epsilon}^{n} \cap\left\{x_1=A\left %
 (1-\epsilon\right)\right\}} 
dx_{2}^{{\alpha}^+}\cdots dx_{n}^{{\alpha}^+}%
\,A\left (1-\epsilon\right )g_{{\alpha}^+}^{1/2}\left (x^{{\alpha}^+}%
\right ){\left |{\Psi}_{{\alpha}^+}\left (x^{{\alpha}^+}\right )%
\right |}^2\\
+\sum_{{\alpha}^-}\int\limits_{ {\tilde M}_{\epsilon}^{n}%
\cap\left\{x_1=\epsilon A \right\}}
dx_{2}%
^{{\alpha}^-}\cdots dx_{n}^{{\alpha}^-}\,\epsilon Ag_{{\alpha}^-}^{1/2}\left %
 (x^{{\alpha}^-}\right ){\left |{\Psi}_{{\alpha}^-}\left %
(x^{{\alpha}^-}\right )\right |}^2\xrightarrow[\epsilon\to 0]{ }\\
\xrightarrow[\epsilon\to 0]{ }1-\int d\mu\,\delta(x_1)A{|\Psi|}^2,
\end{multline*}
 where $d\mu\delta(x_1)$ is a measure at the hypersurface $x_1=0$ induced
by the Riemannian structure of ${\tilde M}^n_{\epsilon}$ .

By using (\ref{general}) one can get
the uncertainty relation of the form (\ref{ours}):
 $\Delta s\cdot\Delta p_s\geqslant 0$.
It is saturated by states which in the map
 $\tilde U_{\alpha}$  can be written as
$$\frac{\exp\left (\frac{2i\pi kx_{1}^{\alpha}}{\hbar A}\right )}%
{g_{\alpha}^{1/4}\left (x^{\alpha}\right )}f_{\alpha}\left (x_{2}^{\alpha},%
\ldots,x_{n}^{\alpha}\right )$$ with functions $f_{\alpha}$  providing
 invariance under any change of map.

\vspace*{2ex}
{\large \bf \begin{center} References. \end{center}}
\vspace{2ex}
\begin{enumerate}
\item Dirac P.A.M. Principles of Quantum Mechanics, 4th ed. (Oxford 
University Press, 1982).
\item Prokhorov L.V. Proc. 6th Int. Conf. Path Integrals from peV to TeV.
Eds. R.Casalbuoni, R.Giachetti, V.Tognetti, R.Vaia, P.Verrucchi.
WS, Singapore, 1999. p.249-252.
\item Menskii M.B. Physics-Uspekhi {\bf 170} (6), 631-647 (2000).
      See also  discussion in Physics-Uspekhi {\bf 171} (4), 437-462 (2001).
\item Chisolm E.D. Am. J. Phys. {\bf 69} (3), 368-371 (2001).  
\item Sukhanov A.D. Physics of Particles and Nuclei {\bf 32} (5), 
619-640 (2001).
\item Bonneau G., Faraut J., Valent G. Am. J. Phys. {\bf 69} (3), 322-331 (2001).
\item Magarill L.I., Romanov D.A., Chaplik A.V. ZhETF {\bf 110} (2),
 669-682 (1996).
\item Bagraev N.T. et al. Semiconductors {\bf 34}
 (7), 817-824 (2000).
\item Green M.B., Schwarz J.H., Witten E. Superstring theory, in two volumes 
(Cambridge: Cambridge University Press, 1987).
\item Kolmogorov A.N. Foundations of the Theory of Probability, 2nd English ed.
(New York: Chelsea, 1956).
\item Schroedinger E. Sitzungber. Preuss. Akad. Wiss. Berlin, 296-303 (1930).
\item Trifonov D.A. arXiv:quant-ph/0307137
\item Sternberg S. Lectures on Differential Geometry, 2nd ed. (Chelsea, 1983).
\item Loudon R. The Quantum Theory of Light (Clarendon Press, Oxford, 1983).
\item Vorontsov Yu.I. Physics-Uspekhi {\bf 172} (8), 907-929 (2002).
\item Borisovich Y.G., Bliznyakov N.M., Izrailevich Y.A., Fomenko T.N.
Introduction to Topology, 2nd ed., in Russian (Nauka Fizmatlit, Moscow, 1995).
\item Warner F. Foundations of Differentiable Manifolds and Lie Groups 
(Springer-Verlag, 1996).
\end{enumerate}
\end{document}